# Unconventional charge-spin conversion in Weyl-semimetal WTe$_2$

Bing Zhao[1,2], Bogdan Karpiak[2], Dmitrii Khokhriakov[2], Annika Johansson[3,4], Anamul Md. Hoque[2], Xiaoguang Xu[1], Yong Jiang[1], Ingrid Mertig[3,4], Saroj P. Dash[2,5,*]

[1]*Beijing Advanced Innovation Center for Materials Genome Engineering, School of Materials Science and Engineering, University of Science and Technology Beijing, Beijing 100083, China*
[2]*Department of Microtechnology and Nanoscience, Chalmers University of Technology, SE-41296, Göteborg, Sweden*
[3]*Institute of Physics, Martin Luther University Halle-Wittenberg, 06099 Halle, Germany*
[4]*Max Planck Institute of Microstructure Physics, Weinberg 2, 06120 Halle, Germany*
[5]*Graphene Center, Chalmers University of Technology, SE-41296 Göteborg, Sweden.*

**Abstract**

An outstanding feature of topological quantum materials is their novel spin topology in the electronic band structures with an expected large charge-to-spin conversion efficiency. Here, we report a charge current-induced spin polarization in the type-II Weyl semimetal candidate WTe$_2$ and efficient spin injection and detection in a graphene channel up to room temperature. Contrary to the conventional spin Hall and Rashba-Edelstein effects, our measurements indicate an unconventional charge-to-spin conversion in WTe$_2$, which is primarily forbidden by the crystal symmetry of the system. Such a large spin polarization can be possible in WTe$_2$ due to a reduced crystal symmetry combined with its large spin Berry curvature, spin-orbit interaction with a novel spin-texture of the Fermi states. We demonstrate a robust and practical method for electrical creation and detection of such a spin polarization using both charge-to-spin conversion and its inverse phenomenon and utilized it for efficient spin injection and detection in a graphene channel up to room temperature. These findings open opportunities for utilizing topological Weyl materials as non-magnetic spin sources in all-electrical van der Waals spintronic circuits and for low-power and high-performance non-volatile spintronic technologies.

**Main**

Topological quantum materials have attracted significant attention in condensed matter physics and spintronic technology because of their unique electronic bands with topologically protected spin textures[1]. After the realization of graphene, topological insulators (TIs), and semimetals with Dirac fermions, Weyl semimetals (WSMs) where the electrons behave as Weyl fermions have been discovered[2]. The WSMs constitute topologically secured Weyl nodes, which exist with opposite chirality in bulk with linear band dispersions in three-dimensional momentum space forming the Weyl cones[3]. The fascinating revelation in a WSM is the presence of nontrivial Fermi-arc surface states that connect the projections of Weyl nodes on the surface Brillouin zone. In a recent breakthrough, WSMs of type-I and -II are realized in TaAs and $WTe_2$ family of materials with symmetric and tilted Weyl cones, respectively[4,5].

$WTe_2$ hosts unique transport phenomena such as chiral anomaly[6,7], unconventional quantum oscillations[8], colossal magnetoresistance[9], spin-orbit torque[10,11], substantial spin Hall effect[12] and quantum spin Hall states in monolayers[13], which opens a new era for physics experiments. Most importantly, novel spin textures have been discovered in WSMs by photoemission experiments, showing spin polarization of Fermi pockets in bulk bands and Fermi arc surface states[14,15]. In such WSMs, the application of an electric field is expected to induce a macroscopic spin polarization, known as the Edelstein effect[16], that can be utilized to generate and detect spin currents efficiently. Moreover, in crystals with lower or broken symmetry compared to conventional metals[17], unconventional spin conductivity components can be existent[18]. Additionally, in the search for spin-polarized current sources in topological quantum materials, various experiments have been reported on TIs[19]. However, a reliable nonlocal measurement for spin polarization in TIs and its utilization for spin injection into non-magnetic materials are so far limited to cryogenic temperatures (below 20K)[20,21], because of the interference from nontrivial bulk bands[19]. Therefore, finding a highly efficient spin-polarized topological material at room temperature is indispensable for practical applications in spintronics and quantum technologies.

Here, we report a highly efficient and unconventional charge-to-spin conversion (CSC) and its inverse phenomena (ICSC), in the type-II Weyl semimetal candidate $WTe_2$ up to room temperature. Importantly, the detected spin polarization in $WTe_2$ is found to be different from the conventional spin Hall effect (SHE) and 2D Rashba-Edelstein effect (REE)[12]. Furthermore, the detection of both the unconventional CSC and ICSC prove the robustness of spin polarization in $WTe_2$ obeying Onsager reciprocity relation and provides a new method for utilization of spin current in graphene for an all-electrical van der Waals spintronic device at room temperature.

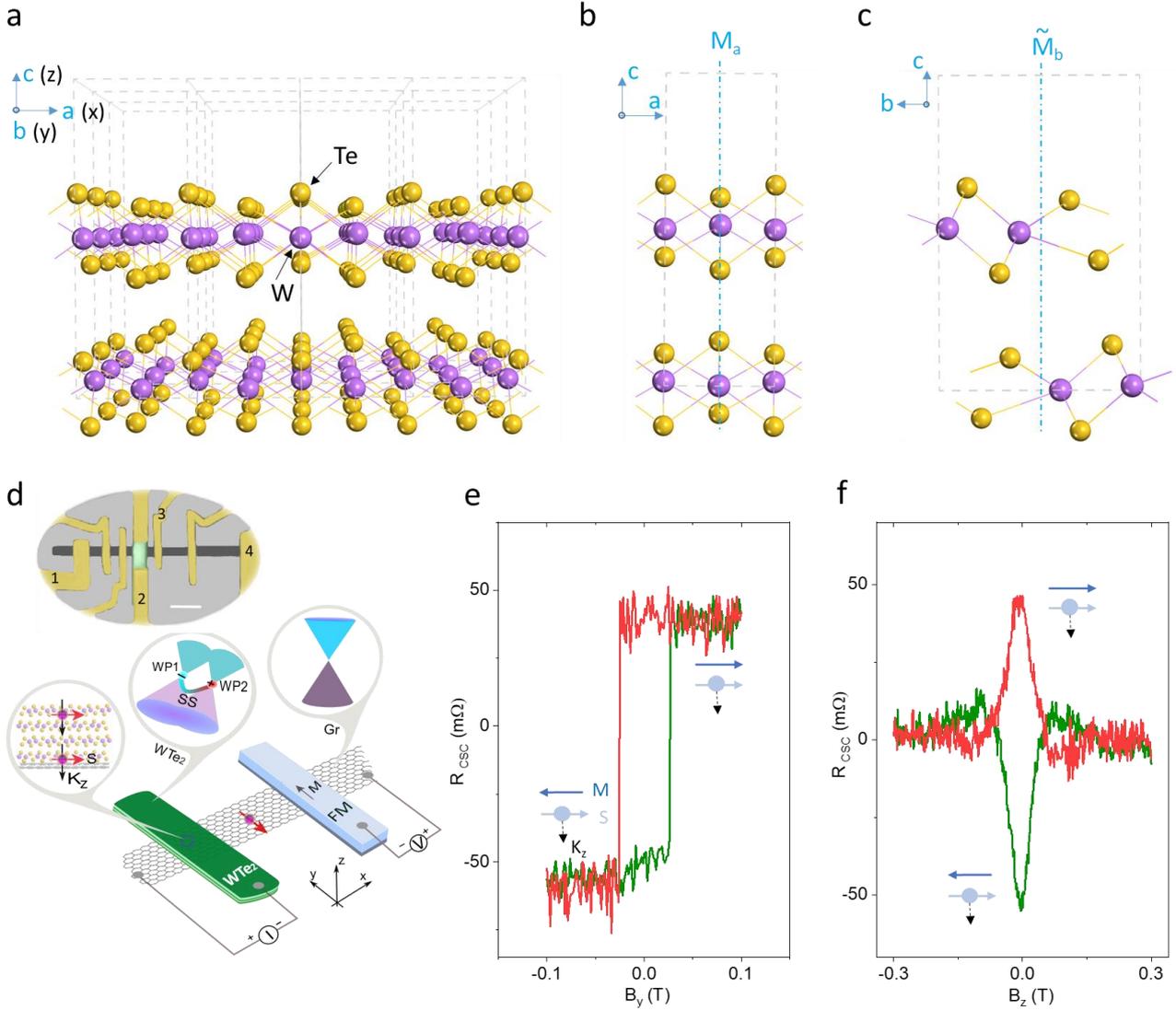

***Figure 1**. **Unconventional charge-to-spin conversion in WTe₂ and efficient spin injection into graphene at room temperature**. **a-c**. Crystal structure of $T_d$ phase bulk WTe$_2$ with a mirror plane $M_a$ (blue dot-dash line) and a glide mirror plane $\widetilde{M_b}$ with the translation of **(a+c)**/2 in the unit cell. **d.** Schematic of measurement geometry and colored device picture (the scale bar is 5 μm) for electrical detection of unconventional charge-spin conversion in WTe$_2$. The investigated device structure consists of a flake of WTe$_2$ (green) with a graphene channel (gray). The ferromagnetic tunnel contact (yellow) on graphene is used to detect the current-induced spin polarization of WTe$_2$. The insets in the schematics show the spin polarization due to perpendicular current component $K_z$, the simplified type-II Weyl semimetal band structure, and the Dirac band diagram of graphene. The upper inset is the optical microscope image of the hybrid WTe$_2$/graphene van der Waals heterostructure device with a ferromagnetic Co detector contact. The contacts (1234) are used as $I_{12(34)}$ and $V_{34(12)}$ for the (I)CSC measurements. **e, f.** The nonlocal spin-valve measurement ($R_{CSC}=V_{CSC}/I$, I is the bias current across WTe$_2$-graphene junction) and corresponding Hanle spin precession signal observed for parallel and anti-parallel orientation of the injected spin (s) from WTe$_2$ and magnetization of ferromagnet (M) with positive and negative magnetic field B sweep directions at I = +50 μA and 300 K in Device 1.*

The measurements of unconventional charge-spin conversion phenomena have been possible by employing a hybrid device structure of $WTe_2$ with graphene channel and ferromagnetic tunnel contacts (FM) in a reliable nonlocal (NL) device geometry. We fabricated van der Waals heterostructures of $WTe_2$ with graphene taking advantage of its layered structure. The schematics and the nanofabricated device picture are shown in Fig. 1d, which consists of $WTe_2$-graphene heterostructure with ferromagnetic tunnel contacts (Co/$TiO_2$) (see Methods for details). The graphene (CVD monolayer[22] and exfoliated few layers) and the $WTe_2$ (20-70 nm in thickness) are exfoliated from single crystals (from Hq Graphene). The Raman spectroscopy characterization shows the $T_d$-bulk phase of $WTe_2$ at room temperature (see Supplementary Fig. S1a). The crystal structure of $WTe_2$ has a nonsymmorphic symmetry, with the space group $Pmn2_1$ for bulk $WTe_2$ crystals with only one mirror plane $M_a$ (bc plane), a glide mirror plane $\widetilde{M}_b$ (ac plane with translation of (**a+c**)/2), and a screw axis $\parallel$ c (Fig.1a, 1b, 1c). Neither two-fold rotational invariance nor inversion symmetry is present in this system[17]. In the measured devices, the heterostructures of $WTe_2$ with graphene show good contact properties with interface resistance in the range of 1-3 kΩ (see Supplementary Fig. S1b), which is an order of magnitude higher than used for detection of conventional CSC in $WTe_2$[12]. The standard spin injection and detection behavior of ferromagnetic tunnel contacts and spin transport properties of graphene was confirmed, as shown in Supplementary Fig. S2.

For the measurement of the unconventional charge-spin conversion effects, an electric current is applied vertically through the $WTe_2$ flake, which generates and injects a spin current into the graphene channel (Fig. 1d). To be noted, here, we used a bias current across the $WTe_2$-graphene junction, which is different from the conventional CSC measurement configuration reported earlier[12] (see details in Supplementary Note 1). The injected spin polarization from $WTe_2$ is detected by a ferromagnetic contact (FM) after transport in a graphene channel by a NL measurement method. Figure 1d shows the NL spin-valve resistance ($R_{CSC}=V_{CSC}/I$) for bias current of I = +50 μA with an in-plane magnetic field ($B_y$) sweep at room temperature. The spin resistance $R_{CSC}$ changes upon reversing the magnetization M direction of the FM detector with respect to the directions of the injected spins (s) from the $WTe_2$ (Fig. 1e).

To prove that the origin of the signal is purely due to a spin current, the Hanle measurements were performed with a perpendicular magnetic field $B_z$ sweep along the z-axis (Fig. 1f). The Hanle effect induces precession of the spins injected from $WTe_2$ and transported in the graphene channel about the $B_z$ field with Larmor frequency of $\omega_L = g\mu_B B_z/\hbar$ (where g is the Landé factor=2, and $\mu_B$ is the Bohr magneton) as the projection of the spin current onto the magnetization of the detector ferromagnet change. The spin injection signal from $WTe_2$ into graphene is reproducibly observed in several devices (six devices were investigated) consisting of both monolayer and few-layer graphene channel with 20−70 nm thick $WTe_2$ flakes (Supplementary Fig. S3 and S4). From these devices, we extracted spin parameters (with spin diffusion length 0.8−2.4μm, spin lifetime in the range 100−400ps) by fitting the Hanle signals (see Supplementary Table 1). The observation of both the spin valve and Hanle signal provide the direct and unambiguous evidence of the creation of current-induced spin polarization in $WTe_2$ and subsequent spin current injection and transport in the graphene channel at room temperature.

In a $WTe_2$-graphene hybrid device, the source of the spin polarization can have several origins, such as spin Hall effect (SHE) and Edelstein effect (EE) from the bulk $WTe_2$, Rashba-Edelstein effect (REE) from the surface states[12], and proximity-induced SHE and REE in graphene[23–25]. Moreover, some of these effects can induce the in-plane spin polarization and can be entangled with each other[12]. To distinguish different sources of the spin polarization, and to identify the origin of the induced spin current in our $WTe_2$ device, control experiments with geometrical dependence were performed.

We first examine the bias current direction dependence of the unconventional CSC signal (Fig. 2a), where the direction of the generated spins *s* in $WTe_2$ is found to be dependent on the polarity of the applied current bias. Reversing the bias current direction ($I_{dc}$=+/-50 µA) in $WTe_2$ results in an opposite spin polarization and hence an inverted hysteretic behavior of the measured spin-valve signal. These measurements show that the direction of spin polarization can be controlled by electrical means, and the spin density is observed to scale linearly with the applied bias current (see Supplementary Fig. S3). The observation of linear bias dependence and a sign reversal behavior with bias current directions rule out the thermal contributions in the measured signal[26,27] (see detail discussion in Note 4).

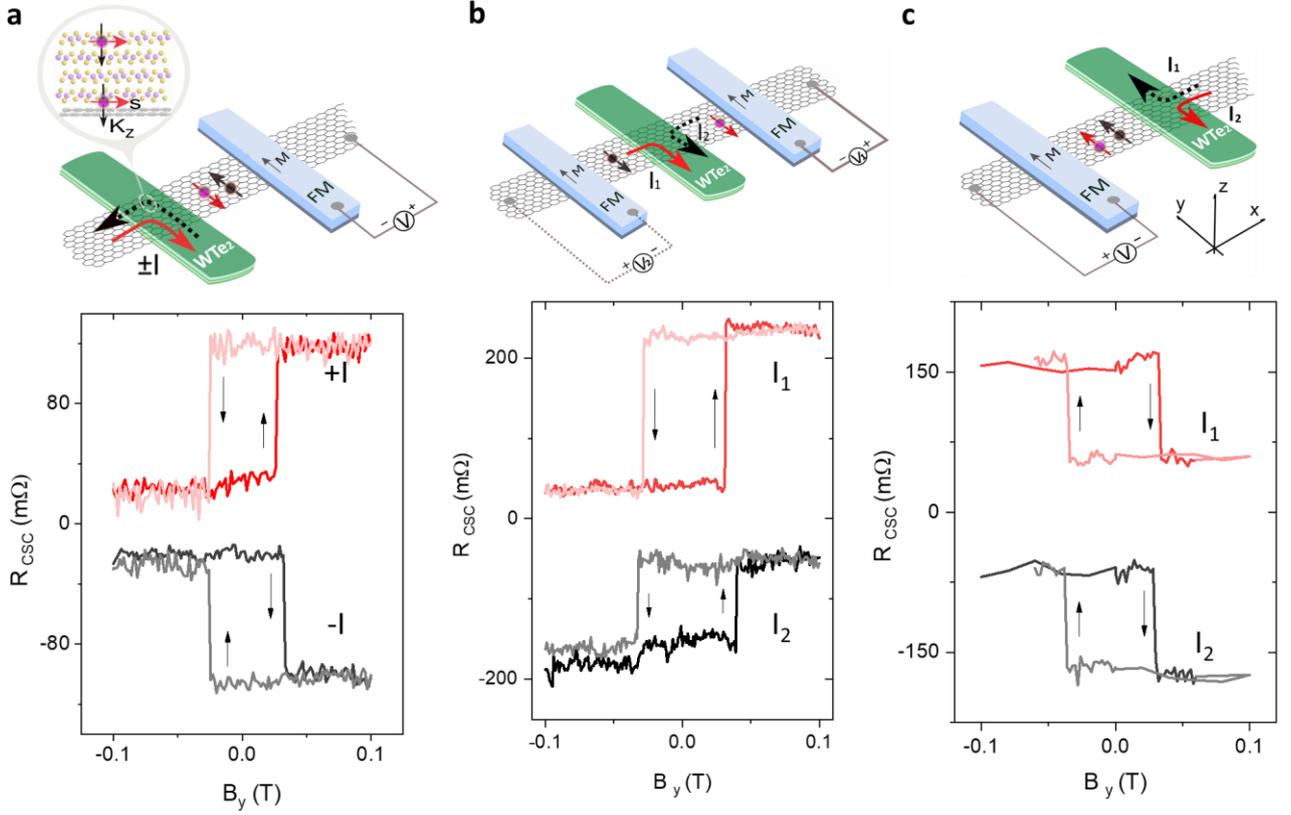

*Figure 2. Geometrical dependence of the unconventional charge-spin conversion effect in $WTe_2$. **a.** Schematics of the measurement geometry (solid and dash arrows show the applied bias current directions) and the corresponding CSC signal with reversal of bias current polarity in Dev 1. The arrows show the switching direction of the signals. The inset shows the spin polarization due to an out-of-plane charge current component $j_c$. **b.** Schematics of the measurement with bias current applied at both sides of the $WTe_2$ flake and corresponding spin valve signals measured in Dev 2. This measurement creates a reversal of the component $K_x$ of the bias current. **c.** Schematics of the measurement with bias current applied at both terminals of the $WTe_2$ flake and the corresponding spin valve signals measured in Dev 3. This measurement creates a reversal of the component $K_y$ of the bias current. A shift in Y-axis is added for the sake of clarity.*

The charge current (I) applied in $WTe_2$ can have three components, i.e., $K_x$, $K_y$, and $K_z$, that can possibly induce the spin polarization (see schematics in Supplementary Fig. S5). We performed control experiments by reversing the bias current polarity along different directions to check the polarity of unconventional CSC signals. As shown in Fig. 2b, the switching directions of unconventional CSC signal ($R_{CSC}$) remain the same with the reversal of bias currents along both sides of the $WTe_2$ flake, i.e., the $K_x$ and -$K_x$ directions. Therefore, the bias current component in $K_x$ direction is not the source for the current-induced spin polarization. Consequently, we can rule out the origins of spin polarizations entangled to $K_x$-direction, like SHE and REE effect in $WTe_2$[12]; and also proximity-induced SHE and REE in graphene[24,28–31]. Secondly, when the bias current was applied either to one or the opposite terminal ($K_y$ and -$K_y$ directions) of $WTe_2$ (Fig. 2c), the $R_{csc}$ signal switching directions remain the same. This rules out the contribution to the spin polarization from the current component in $K_y$ direction. Both the control experiments in $K_x$ and $K_y$ directions were reproduced in other batches of devices (see Supplementary Fig. S6 and Fig. S7). To be noted, the switching direction of the spin-switch signal is observed to be different for different devices (Dev 2 and 3 in Fig. 2b and 2c). This can be due to the uncertainty in different crystal orientations (±a, ±b axis) of the exfoliated WTe$_2$ flakes relative to the detector FM.

All these control experiments indicate that the current component in $K_z$ direction in WTe$_2$ is the primary source for the generation of spin polarization in the measurement geometry of WTe$_2$-graphene devices. Therefore, the origin of the spin signal can be attributed to the out-of-plane current-induced spin momentum locking of the spin-polarized Fermi states at room temperature[14,15]. Moreover, the out-of-plane current component along $K_z$ can have two contributions: one ($K_{z,B}$) through the bulk states; and the other ($K_{z,S}$) via the surface states (see Supplementary Fig. 5c). From the recent ARPES measurement on WTe$_2$[32], it is clear that the surface states barely disperse with $k_z$ over an entire Brillouin zone and can thus be considered fully two-dimensional (2D) in the $k_x$-$k_z$ plane, i.e. the Fermi lines are straight lines at the edge of the bulk electron and hole pockets. Therefore, the current in $K_z$ direction is not accompanied by any significant transport originating from the surface states: nearly no surface states would contribute to the observed unconventional CSC signal (see details in Supplementary Note 5). Moreover, according to our control experiments (Fig. 2), we know that it is the current along z-axis that induces the spin polarization **s**, which are parallel to FM magnetic moments M, i. e. **s**//**M**//**y**. We also know that the spin current **j$_s$** is along the **z**-axis in our measurement geometry, i.e., **j$_s$**//**z**//**K$_z$**, which is also the direction of the charge current. However, in the conventional SHE measurements, charge current, spin current, and spin polarization should be mutually perpendicular, following the right-hand rule, i.e., **j$_s$**⊥**K$_z$**⊥**s**[12,33]. Therefore, the measured data in the present experimental configuration do not follow the conventional SHE rules (see Supplementary Note 1). Further, the unconventional CSC signal in this geometry can also neither originate from an unconventional spin conductivity nor the Edelstein effect of the spin-polarized Fermi states for symmetry reasons, as long as the symmetry operations of the space group Pmn2$_1$, in particular mirror and glide mirror symmetries, are present (see Supplementary Note 5). However, if the symmetry of WTe$_2$ is reduced, e.g., by strain[34–36], magnetic field[37,38] or the interfaces [39] between WTe$_2$ and graphene, both unconventional spin Hall effect and Edelstein effect can give rise to the observed spin polarization (see details in Supplementary material Note 5). However, our observation suggests that a magnetic field is less likely to be the origin of the unconventional CSC signal (see details in Supplementary material Note 3). Strain can arise in the WTe$_2$ flakes due to fabrication of contacts and interfaces with different materials in the device, breaking both mirror M$_a$ and glide mirror $\widetilde{M}_b$ symmetry. Furthermore, the interface between WTe$_2$ and graphene can as well break glide mirror and screw symmetries locally. This symmetry breaking leads to the occurrence of a y-polarized spin current and (or) a homogeneous spin density in y direction, leading to a spin current **j$_s$** || **K$_z$** (here we name it $j_s^y$). Thus, although the symmetries of space group Pmn2$_1$ prohibit the generation of a spin

current $j_s^y$ by the spin Hall and Edelstein effects, breaking the crystal symmetries by strain or the occurring interface allows the emergence of the observed spin current.

To further confirm the orientation of spin polarization from WTe$_2$, we also performed angle-dependent measurements of the unconventional CSC signals both with in-plane and out-of-plane B field sweeps in Dev 4 (see Supplementary Fig. S8). The measured unconventional CSC signal is observed to evolve from a step-like spin valve signal to a Hanle signal with changing the angle from 0 to 90 deg. Considering the direction of the spin current j$_s$, the spin polarization **s,** and bias current **K$_z$**, the contribution of the conventional CSC from WTe$_2$ can also be ruled out. These systematic measurements again support the observation of the unconventional CSC in bulk WTe$_2$. The unconventional CSC measurements in WTe$_2$ using both the spin valve and Hanle geometry were also performed as a function of gate voltage (V$_g$) in Dev 4 (see Supplementary Fig. S9), where an enhancement of the unconventional CSC signal magnitude is observed close to the Dirac point of graphene. As the metallic WTe$_2$ channel resistance does not show any noticeable modulation with V$_g$ (Supplementary Fig. S10), the increase of unconventional CSC signal can be attributed to the increase in the graphene channel resistance in the heterostructure and conductivity matching issues at the interface (Supplementary Fig. S9c). However, in the conventional spin valve signals with both FM injector and detector contacts on graphene, a small modulation of spin signal magnitude is observed (Supplementary Fig. S10)[40]. This increasing trend of CSC signal with V$_g$ suggests a possible enhancement of the spin injection efficiency from WTe$_2$ to graphene near the graphene Dirac point.

In comparison to the symmetric Hanle curves in other devices, Device 2 shows an in-plane asymmetric characteristic in the Hanle spin precession signal (Fig. 3a). By decomposing the measured raw data mathematically, we obtained the symmetric and the anti-symmetric spin components, i.e., the spins along FM and the spins perpendicular to FM, respectively. Specifically, the magnitude of the symmetric (R$_1$) and the anti-symmetric (R$_2$) components are extracted by fitting the Hanle curves. Thus, the out-of-plane charge current generates a net spin polarization and/or a spin current, which is polarized with an angle of 76.4° (=arctan(R$_1$/R$_2$)) with respect to the FM. Such an asymmetric Hanle curve can be caused by the coexistence of spin polarization components parallel and perpendicular to the ferromagnet[41]. This can be explained by considering the simultaneously broken mirror symmetry M$_a$ and glide symmetry $\widetilde{M}_b$ (see the symmetry analysis in Supplementary Note 5). We cannot distinguish the orientation of the samples in the present case, however, considering the randomly distributed a(b)-axis in the exfoliated WTe$_2$ flakes relative to the crystal boundary and presence of strain as well as the glide-symmetry ($\widetilde{M}_b$) -breaking interface in the devices, such an unconventional and asymmetric signal due to charge-to-spin convention is observed.

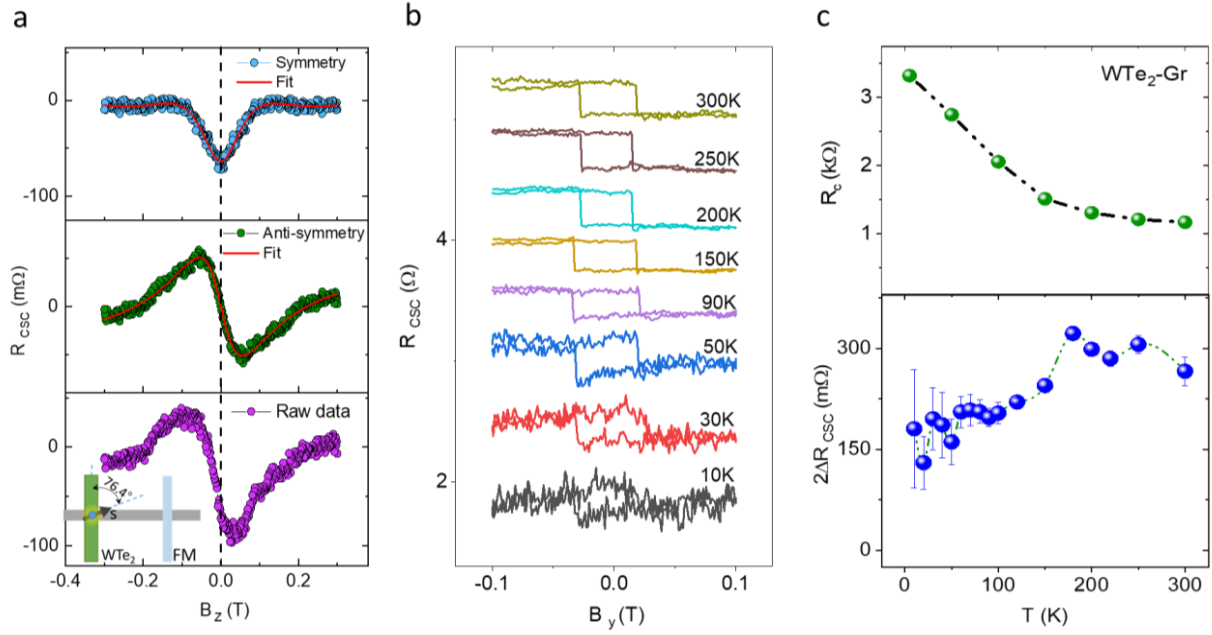

*Figure 3. Observation of the in-plane asymmetric Hanle signal and temperature dependence of the unconventional charge-spin conversion in WTe$_2$. **a.** The measured asymmetric Hanle raw data (bottom panel) of the unconventional CSC signal in Device 2. The symmetric (top panel) and anti-symmetric (middle panel) components of the Hanle signal and the corresponding fittings. The inset shows the measurement geometry with charge current-induced spin polarization **s** in WTe$_2$ can be with an angle relative to the detector FM. **b.** Temperature dependence spin-switch signal of the unconventional CSC effect in WTe$_2$ for Device 2 ($R_{CSC}=V_{CSC}/I$, with I=-50µA in the range of 10-300 K. **c.** Top panel: Temperature dependence of the WTe$_2$-graphene interface resistance. Bottom panel: Temperature dependence of the CSC signal magnitude. The error bars are estimated from the noise in the measured CSC signal.*

Next, the temperature-dependent measurements of the unconventional CSC in WTe$_2$ were carried out to correlate the basic characteristics of the Weyl materials and their spin polarization. The WTe$_2$ has been experimentally verified to be a type-II Weyl semimetal by observing the negative magnetoresistance[8,42] and the anomalous-quantum oscillation[8] at low temperatures. However, the verification of WTe$_2$ to be a type-II Weyl semimetal at room temperature is still under debate[32,43,44], because the momentum difference for Weyl points (WPs) is beyond the resolution of ARPES measurement[32,45],[7]. Hence, it is intriguing to check the temperature dependence of the signal and its relationship with the possible Weyl phase transition at a lower temperature[8,42]. Figure 3a shows the temperature dependence of the unconventional CSC signals from WTe$_2$, measured in Dev 2 in the temperature range of 10 – 300 K at a fixed bias current I =-50µA. The switching direction of the measured unconventional CSC signal in WTe$_2$ remains the same throughout the temperature range, indicating that the origin of the spin polarization remains the same. The apparent slightly larger switching field observed for the signal at lower temperatures is due to an increase in the coercive field of the FM detector contact. The modulation of the magnitude of the unconventional CSC signal of WTe$_2$ is plotted with temperature in Fig. 3b (lower panel), where two different regimes were observed. The signal magnitude is weakly temperature-dependent in the range 175 – 300K, whereas a notable decrease is observed below 175K. As spin transport parameters in graphene are known to weakly dependent on temperature[46], we identified that the decrease in unconventional CSC magnitude could be due to the increase in WTe$_2$-graphene contact resistance at lower temperatures (Fig. 3b upper panel). The increase in WTe$_2$-graphene contact resistance could cause a decrease in spin injection efficiency form WTe$_2$ into graphene, while the other parameters were very stable with temperature (see Supplementary Fig. S11). The increased noise level in the unconventional CSC signal at the lower temperature can also be due to the larger WTe$_2$-graphene contact resistances. All these measurements suggest that the observed unconventional CSC effect is robust from 10K to room temperature, and the spin-polarized bulk Fermi states in WTe$_2$ should be present through all the measured temperature range.

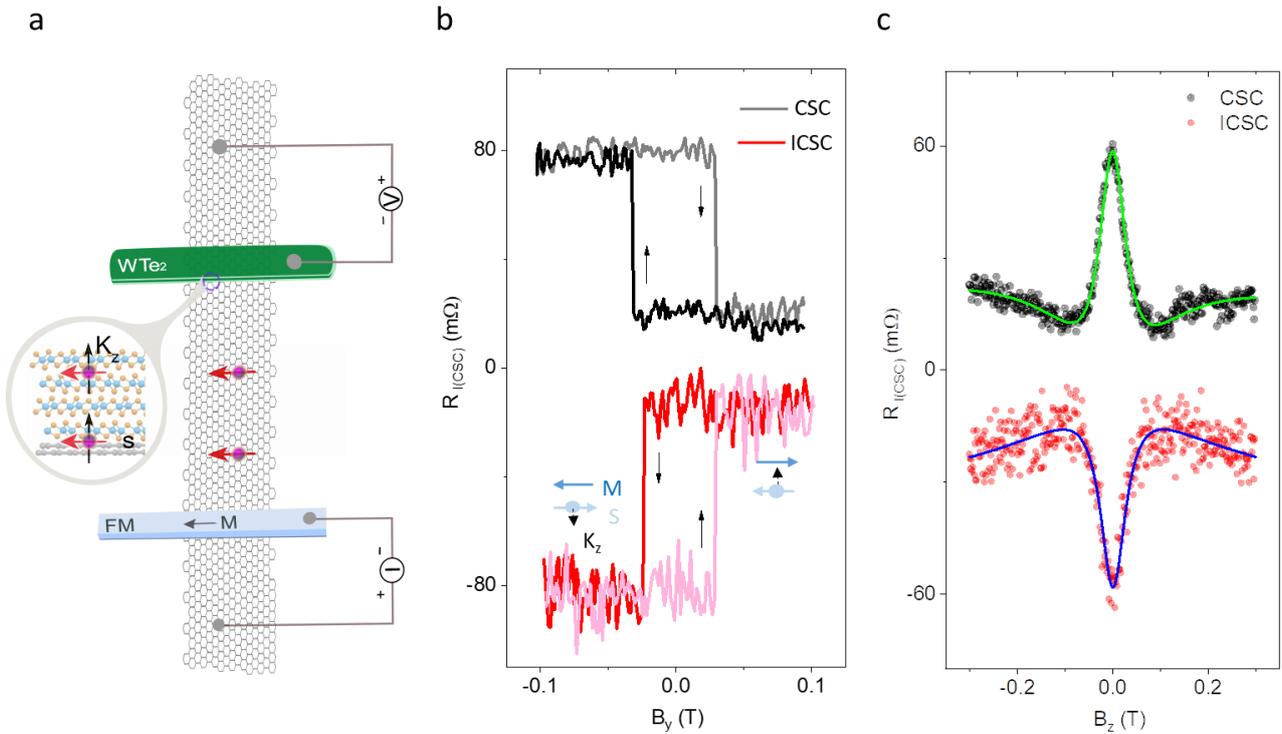

*Figure 4: Inverse charge-spin conversion in WTe$_2$ at room temperature. a. Schematics of the inverse charge-spin conversion (ICSC) measurement configuration with spin current injected into the WTe$_2$ from the FM/graphene structure. b, c. Measured data of both spin valve and corresponding z Hanle signal with the fitting curve for both CSC and ICSC in Device 1 with an application of I=-70µA at room temperature. The Hanle signal of the (I)CSC is defined by $R_{(I)CSC}= (V_{(I)CSC}(P)-V_{(I)CSC}(AP))/(2*I)$.*

To check the reciprocity[47] of the unconventional charge-spin conversion effect in WTe$_2$, the inverse charge-spin conversion (ICSC)[48,49] measurements were also performed at room temperature (see Fig. 4a). Here, the spin current is injected from an FM tunnel contact into a graphene channel and subsequently detected by the WTe$_2$ due to the ICSC effect in a NL measurement geometry. The spin current with spin polarization **s** along the y-axis is absorbed into the WTe$_2$, resulting in a net charge current in the **K$_z$** direction due to the inverse charge-spin conversion. By reversing the FM magnetic moments by an external in-plane magnetic field sweep, the opposite spins **-s** are injected into the graphene channel and subsequently absorbed by the WTe$_2$, which induces a charge current **-K$_z$**. An apparent hysteric switching behavior of the measured voltage signal is observed with B$_y$ magnetic field sweep, with the switching fields corresponding to the magnetization switching of the spin injector FM electrode (Fig. 4b). Importantly, both the ICSC and CSC signals measured in the same device show a good Onsager reciprocity with comparable signal magnitude and opposite switching directions, i.e., R$_{CSC}$(B)=R$_{ICSC}$(-B) (Fig. 4b). The ICSC effect is reproducibly observed in different devices and the magnitude of the signal is also found to scale linearly with spin injection bias current and changes sign with the bias polarity (see Supplementary Fig. S12 for data on Dev 2). As a confirmatory test for the observed ICSC effect, corresponding Hanle spin precession measurements were also performed with application of an out-of-plane magnetic field B$_z$ in the same NL measurement geometry. Figure 4c shows the measured modulation of signal R$_{ICSC}$=V$_{ICSC}$/I as detected by ICSC in WTe$_2$ due to spin diffusion and precession in the graphene channel.

We estimate the unconventional CSC efficiency α$_{CSC}$ (=j$_s$/j$_c$) in WTe$_2$ to be up to 9% at room temperature by fitting the Hanle curves of (I)CSC signals (see details in Supplementary Note 2). The calculated lower limit of Edelstein length is λ$_{EE}$ = α$_{CSC}$λ$_{WTe_2}$ ≈ 0.72nm (considering spin diffusion length of WTe$_2$ λ$_{WTe_2}$=8nm[12]). The observed large charge-spin conversion efficiency in WTe$_2$ is believed to be due to the spin polarization of Fermi states, broken space inversion symmetry, and a significant influence of SOI in WTe$_2$ as known from the band structure calculations and spin-resolved ARPES results[15]. Interestingly, our measurements show that the charge-spin conversion is not restricted to the 2D surface but originates in the bulk Weyl semimetal WTe$_2$. From our control experiments, we could rule out the origins of the observed spin polarization related to conventional SHE and REE in WTe$_2$[12], and also proximity induced SHE and REE in graphene[24,28–31]. These observations of the unconventional CSC are fundamentally different from the conventional REE, which is an interface phenomenon where the spins and current density are confined in the 2D plane[49,50] as measured in the heterostructures of metals[49] and oxides[51], topological insulator[52,53], Transition metal dichalcogenides (TMDCs)[54] and in graphene heterostructures with MoS$_2$[28], WS$_2$[29,31], TaS$_2$[25], MoTe$_2$[24] and (Bi$_{0.15}$Sb$_{0.85}$)$_2$Te$_3$[55]. This unconventional charge-spin conversion phenomenon in WTe$_2$, however, is shown to be useful for injection and detection of spin polarization in graphene at room temperature, avoiding problems existing in topological insulators[20,21], and open ways for spintronic devices without the use of traditional ferromagnets. The WTe$_2$ based van der Waals heterostructure devices also provide the advantage that their operating temperature is not restricted by a Curie temperature (T$_c$), such as recently discovered 2D ferromagnets have T$_c$ much below the room temperature [56]. Other advantages are that the direction of spin polarization in WTe$_2$ can be controlled by using an electric bias current, instead of using an external magnetic field in case of FMs to switch the magnetic moments. From the application point of view, spin-orbit torque (SOT) studies[10,11] show that WTe$_2$ can be more efficient and energy-saving in SOT technologies compared to traditional heavy metals.

In summary, we demonstrated the electrical creation, detection, and control of the unconventional charge-spin conversion and its inverse phenomenon in type-II Weyl semimetal candidate WTe$_2$ up to room temperature. Contrary to conventional bulk spin Hall effect and surface states dominated Rashba-Edelstein effect, the charge-spin conversion is shown here to be created in WTe$_2$ due to unconventional spin Hall effect or (and) Edelstein effect. The unconventional spin conductivities in WTe$_2$ are allowed by considering strain as well as the WTe$_2$/graphene interface to break the crystal symmetry. The spin polarization created in WTe$_2$ is shown to be utilized for spin injection and detection in a graphene channel in an all-electrical van der Waals heterostructure spintronic device at room temperature, which circumvents the problem existing in topological insulators for spin injection into graphene below 20K[20,21]. Such unique spin-polarized electronic states in Weyl semimetal candidates with novel spin topologies can be further tuned by tailoring their electronic band structure through enhancing their spin-orbit interaction strength, increasing the separation between the Weyl nodes through Berry curvature design, and controlling strain to break the crystal symmetry. These findings in Weyl semimetal WTe$_2$ for efficiently transforming the electric current into a spin polarization at room temperature is highly desirable for energy-efficient spintronic memory and information processing technologies[57].

**Note**

After submission of our manuscript, we noticed two very recent papers on the multidirectional and unconventional charge-spin conversion in MoTe$_2$ [36,39]. Our results on Weyl semimetal candidate WTe$_2$ show an efficient and unconventional charge-spin conversion, which is different from conventional SHE and REE and demonstrates a practical approach for efficient generation and injection of spin polarization into graphene channel up to room temperature.

**Methods**

The devices with monolayer CVD graphene on Si/SiO$_2$ substrate (from Groltex) were patterned by electron beam lithography (EBL) followed by an O$_2$ plasma etching. The WTe$_2$ (from Hq Graphene) flakes were exfoliated and dry transferred on to the CVD graphene in the N$_2$ atmosphere inside a glovebox. The few layers graphene devices were mechanically transferred onto the n-doped Si substrate with 300 nm SiO$_2$. The WTe$_2$ flakes were exfoliated and dry transferred on to the few-layer graphene in by a transfer stage in a class 100 cleanroom environment. For the preparation of ferromagnetic tunnel contacts to graphene, a two-step deposition and oxidation process was adopted, 0.4nm Ti was deposited, followed by a 30 Torr O$_2$ oxidation for 10 mins each, followed by 100 nm Co deposition. Measurements were performed inside a vacuum cryostat and a PPMS measurement system in the temperature range of 10-300K with a magnetic field and a sample rotation stage. The electronic measurements were carried out using current source Keithley 6221, nanometer 2182A, and dual-channel source meter Keithley 2612B.

**Data availability**

The data that support the findings of this study are available from the corresponding authors on reasonable request.

**Acknowledgments**

The authors at Chalmers University of Technology, Sweden acknowledge financial supports from EU Graphene Flagship (Core 1 No. 604391, Core 2 No. 785219, and Core 3 No. 881603), Swedish Research Council VR project grants (No. 2016-03658), EU FlagEra project (from Swedish Research council VR No. 2015-06813), Graphene center and the AoA Nano program at Chalmers University of Technology. The authors from the University of Science and Technology, Beijing, China, acknowledge financial supports from the National Basic Research Program of China (Grant No. 2015CB921502) and the National Natural Science Foundation of China (Grant Nos. 51731003, 51471029). Bing Zhao would like to thank the financial support from the program of China Scholarships Council (File No. 201706460036) for his two years of research at Chalmers. The authors from Martin Luther University Halle-Wittenberg acknowledge support by CRC/TRR 227 of Deutsche Forschungsgemeinschaft (DFG). The authors would like to thank Binghai Yan, Marcos Guimaraes, C. K. Safeer and Felix Casanova for useful discussions.

**Author information**

**Contributions**

SPD and BZ conceived the idea and designed the experiments. BZ, BK, DK, AMH, SPD fabricated and measured the devices at the Chalmers University of Technology. BZ and SPD analyzed, interpreted the experimental data, compiled the figures, and wrote the manuscript. AJ and IM performed theoretical calculations. BK, DK, AMH, XX, YJ discussed the results and provided feedback on the manuscript. SPD supervised the research.

**Competing interests**

The authors declare no competing financial interests.

**Corresponding authors:**

Saroj P. Dash, Email: saroj.dash@chalmers.se